**Editorial:**
**Making GIScience Research More Open Access**

Bin Jiang
Department of Technology and Built Environment, Division of Geomatics
University of Gävle, SE-801 76 Gävle, Sweden
Email: bin.jiang@hig.se

This special issue is special not just because of the emerging theme of data-intensive geospatial computing, and but also in the sense that all the data and source codes used in the papers are made freely available. In the initial call for papers (http://fromto.hig.se/~bjg/ijgis/), authors were encouraged to consider the option of support materials for archiving data and video clips. With the endorsement of the editor in chief, Dr. Brian Lees, a short note was inserted in the decision letter following the first round of review comments. The short note reads as follows:

> *To enhance scholarly communication and as a distinguishing feature of data-intensive science, we ask authors to submit all test data and algorithmic source codes as Supplementary Materials. These will be made accessible from the online version of your paper. For information about this, see paragraph 6 of our Instructions To Authors at http://www.tandf.co.uk/journals/journal.asp?issn=1365-8816&linktype=44*
>
> *We encourage all authors to package your data and source codes in such a way that the readers of IJGIS can replicate your experiments by simply clicking some buttons to replicate all your results. If you were unable to archive the test data and the source codes for whatever reasons, we would not accept your paper in this special issue. In this case, we would recommend your paper to IJGIS regular issues should your paper have passed the review process.*

Collected in this special issue is a set of papers that deal with various issues on data-intensive geospatial computing, and importantly whose authors have published their data and source codes. All data and codes are made available either at the publisher's website, with links to the individual papers, or at a complementary website (http://fromto.hig.se/~bjg/ijgis/supplementary/) for large scale data that are too large to archive at the publisher's site. We believe that making this information freely available is an important aspect of data-intensive science (Hey et al. 2009). Even with the publication of data, we are still far from the vision of the late Jim Gray of Microsoft Research, who envisioned *"a world of scholarly resources—text, databases, and any other associated materials—that were seamlessly navigable and interoperable."* This editorial elaborates on scholarly communication, with particular attention to publishing data alongside papers and the emergence of open access journals, in order to make our research more open access.

Scholarly communication is one of the two fundamental issues underlying data-intensive computing. The first issue is the cyberinfrastructure, which has been discussed widely among geographic information science (GIScience) researchers (e.g., Wang and Liu 2009, Yang et al. 2010, Yang et al. 2011). In this regard, it should be noted that state of the art personal computers are powerful enough to carry out some data-intensive computing at the scale of gigabytes (Jiang 2011). For example, geographic computation that relied on high performance computing 10 years ago (Openshaw and Turton 2000) can easily be performed using current personal computers. Current research has been exploring gigabytes of OpenStreetMap data (Bennett 2010) to gain insight into the scaling of geographic space or urban structures (Jiang and Liu 2012, Jiang and Jia 2011). We believe that these insights can only be obtained with extensive rather than small geospatial data. Data size does indeed matter! These data are freely obtained, and any resulting data can be freely shared with anyone interested in further exploration of the data for research purposes.

Today, global positioning system (GPS) receivers have become ubiquitous, and one can easily track one's own movement. However, capturing the movements of a large population still remains a luxury. For example, imagine thousands of taxicabs moving around a city, with their locations constantly changing. Every few seconds, there would be a new location captured from a GPS receiver attached to individual taxicabs. Collectively, these locations, which have been acquired on a 24/7 basis, represent an image of the city (Lynch 1960). On the one hand, this image is clearly viewable as an ordinary city map. On the other hand, frequently visited locations are distinguished from the thousands of other locations. The data size is massive and is as large as one gigabyte per day. With such data sizes, one can study individual moving behavior, spatial heterogeneity in terms of individual spatial positions, and even extract underlying street networks using smart algorithms. Both spatial and temporal patterns can be further explored with large amounts of GPS data. In fact, our daily activities, such as emails, mobile phone calls, blogs, and credit card purchases all leave digital traces in one way or another. Data-intensive computing of these digital traces can provide a comprehensive picture of both individual and group behaviors. The capacity for collecting and analyzing massive amounts of data has been significantly transforming traditional social science toward a more data-driven computational social science (Lazer et al. 2009) to better understand individuals and collectives, or society as a whole.

Therefore, publishing data or data sharing is very important and is critical for science. We can see examples of this importance for science historically. For example, if Tycho Brahe had not shared his painstakingly collected data with Johannes Kepler, the story about the discovery of the laws of planetary motion would be different. As a GIScience researcher, I have been frustrated by the fact that source codes and data are rarely shared among those interested. I do understand that there are some restrictions on sharing data and source codes. For example, scientists may have signed nondisclosure agreements with the data providers because of the sensitivity of the information or privacy. Additionally, scientists may spend a great amount of time on data collection and processing, and they therefore may not want to share their data freely. For whatever reason or constraint, this barrier is tremendous for innovative research because, without sharing, results or conclusions cannot be replicated. More importantly, through data sharing, other researchers may expand upon the initial results and conclusions to find something hidden in the data. However, this situation is changing because more and more geospatial data are becoming freely available through resources such as the OpenStreetMap data or volunteered geographic information in general (Goodchild 2007). GIScience researchers could learn from other disciplines, such as genomics and biology, where there is a good tradition or culture for data sharing. Moreover, many publishers or journals have offered options for archiving supplementary materials and provide a link to individual papers published online (e.g., *IJGIS and Computers, Environment and Urban Systems)*.

The notion of open access journals has emerged as a new means for scholarly communication. These journals combine traditional peer review and Web 2.0 technologies for scientific communication. In this regard, the Public Library of Science (PLoS) has a series of open access journals, such as PLoS Biology and PLoS Medicine. PLoS ONE is probably the most successful open access journal, and it is available for all disciplines in science and medicine (MacCallum 2006). Within a few years of its launch in 2006, its impact factor has reached 4.35. In 2010, PLoS ONE published 6,749 articles according to the statistics provided by its editor Peter Binfield, and it has become the world's largest journal in terms of published papers. The open access journals adopt a new philosophy of peer-reviewing and publishing, where papers are published first and judged later (Giles 2007). Acceptance of a paper is not based on how significant the result it, but rather on whether the science in the paper has been performed appropriately. Additional reviewing takes place after the paper has been published. For example, every published article is open for discussion and comments from the scientific community. This process is supported by Web 2.0 technologies. Although open access journals are open in terms of access to papers, data and source codes are not necessarily available. The PLoS ONE's guidelines for authors clearly state that datasets should be submitted as supporting information; however, as far as I know, this is not a mandate. Many published papers in PLoS ONE have not put the datasets online freely for further research. This practice is not consistent with the spirit of data-intensive computing.

Long before the current open access journals movement, scientific community has attempted to provide public platforms for sharing unpublished papers. For example, arXiv (arxiv.org) is an archive for individual researchers to upload their preprints before submitting to scientific journals. It covers all subfields in physics, mathematics, computer science, quantitative biology, quantitative finance and statistics. It has archived over a half million preprints, with a growing rate of 5000 new e-prints per month. In the above subfields, it is commonplace to cite a paper available from the arXiv.

GIScience, or computational geography, is a relatively young discipline and often involves large datasets about our environments and human activities. Data-intensive geospatial computing of massive geographic information can help us to understand the underlying mechanisms of geographic structures and processes. We should be inspired by data-intensive science and data sharing much like our fellow scientists working in genomics and biology. In fact, GIScience researchers have recently launched two open access journals: *Journal of Spatial Information Science* and *ISPRS International Journal of Geo-Information*. Consequently, we foresee that more GIScience articles will be published in the open access manner in the near future.

I am not going to summarize individual contributions collected in this special issue. Instead I would like to make a reminder: for anyone interested in accessing the large amounts of GPS data mentioned earlier, it is provided with this special issue. I would like to express my thanks to Dr. Brian Lees, who trusted me as a guest editor for this special issue. My sincere thanks are extended to the referees for providing valuable comments on individual papers and to the authors who share the data-intensive computing spirit with me and who make all source codes and data available for free.